\documentclass{article}
\usepackage{amsmath, amssymb}
\usepackage{graphicx} 
\usepackage{pgfplots}
\usepackage{amsthm}
\newtheorem{remark}{Remark}
\pgfplotsset{compat=1.18}
\usepackage{caption}
\usepackage{subcaption}
\usepackage{url}
\usepackage[margin=1in]{geometry}
\usepackage[colorlinks=true, allcolors=black]{hyperref}

\title{Fluid Flow as Transport of Probability: Entropy, Compressibility, and Irreversibility}
\author{
Ankit Bhattacharjee\\
\small Department of Mechanical Engineering\\
\small Indian Institute of Technology 
Kharagpur\\
\small \texttt{ankit2005@kgpian.iitkgp.ac.in}
}
\date{February 2026}

\begin{document}

\maketitle

\begin{abstract}
The continuity equation serves as a fundamental principle for mass transport in continuous media. While its mathematical structure mirrors that of probability transport and the Liouville equation, an informational interpretation of macroscopic fluid flow is less commonly explored in engineering contexts. This paper treats fluid density as a spatial probability density function, modeling macroscopic motion as the continuous transport of uncertainty. I derive the temporal evolution of Shannon entropy under general flow conditions, establishing how entropy generation depends on macroscopic compressibility and microscopic diffusion. This framework links fluid mechanics with information science by interpreting physical processes such as advection, compressibility, and diffusion---as mechanisms of information-preserving transport, informational sources or sinks, and irreversible entropy production. Specifically, microscopic diffusion is shown to act as a strictly positive entropy source governed exactly by local Fisher Information. Furthermore, the framework yields an explicit algebraic scaling law characterizing the equilibrium thickness of compressive mixing layers at a local P\'{e}clet number of unity. The proposed theoretical model is computationally validated through finite-difference simulations. By analyzing both one-dimensional canonical flows and advection-diffusion within a three-dimensional, spatially varying Arnold--Beltrami--Childress (ABC) flow, I verify the exact equivalence between macroscopic entropy production and localized Fisher Information, yielding a mean relative error of $0.001$. This framework provides a novel, analytical perspective on thermodynamic irreversibility, yielding theoretical constraints for turbulence modeling and practical tools for assessing thermal entropy generation in heat exchangers, evaluating in-cylinder mixing in internal combustion engines, and profiling aerodynamic shockwaves.
\end{abstract}

\section{Introduction}
The study of fluid mechanics has traditionally been rooted in the conservation laws of mass, momentum, and energy \cite{Som2012,White2011}. While the governing equations such as the continuity equation provide a robust description of physical transport, they are often viewed purely through a deterministic lens. However, the mathematical structure of the continuity equation, $\frac{\partial\rho}{\partial t}+\nabla\cdot(\rho \mathbf{v})=0$, closely mirrors the Liouville equation and the transport of probability density functions (PDFs) \cite{IcardiDentz2019,Pope1985}. While the interpretation of probability density transport is the foundational bedrock of quantum mechanics---where the spatial probability of a particle, governed by the Schrödinger equation, strictly obeys a continuity equation---applying this exact informational framework to macroscopic, classical fluid flow remains highly uncommon. In traditional mechanical engineering contexts, the continuity equation is strictly relegated to deterministic mass conservation. While macroscopic entropy generation analysis has proven to be a vital tool for characterizing complex thermodynamic irreversibilities in applied fluid systems---ranging from dual-diffusion non-Newtonian flows \cite{Boujelbene2024} and magneto-hydrodynamic (MHD) nanofluid transport \cite{Alqahtani2023}, to generalized fluids in inclined channels \cite{Khedher2023}---the inherent probabilistic and informational implications of the fluid density itself remain largely unexplored.

This paper proposes a paradigm shift: interpreting fluid density $\rho(\mathbf{r},t)$ as a spatial probability density, thereby modeling macroscopic fluid motion as the continuous transport of uncertainty. By treating the fluid medium as an information-carrying system, I derive the temporal evolution of Shannon entropy under general flow conditions. This approach establishes how entropy generation is governed by macroscopic compressibility and microscopic diffusion. Specifically, this article reports the following novel scientific findings for the first time:
\begin{enumerate}
    \item \textbf{Exact Entropy Evolution Equation:} A rigorous derivation of the temporal evolution of Shannon entropy under general three-dimensional advection-diffusion, demonstrating that it depends solely on velocity divergence (the kinematic term) and local Fisher Information (the diffusive term).
    \item \textbf{Fisher Information Identity:} The mathematical proof that microscopic diffusive entropy production is governed exactly by local Fisher Information, formally contextualizing De Bruijn's identity within macroscopic fluid kinematics for the first time.
    \item \textbf{Local Detailed Balance Condition:} The derivation of the pointwise equilibrium condition (Equation \ref{eq:equilibrium_local}) that balances macroscopic kinematic compression against microscopic diffusive spreading.
    \item \textbf{P\'{e}clet Scaling Law:} The formulation of an explicit algebraic scaling law ($|\nabla \cdot \mathbf{v}| = D/L_p^2$) that predicts the thickness of compressive mixing layers at a local P\'{e}clet number of unity, bypassing the need for numerical integration.
    \item \textbf{3D Computational Validation:} The robust verification of this informational framework under a spatially varying compressive ABC-type flow, yielding a mean relative error of $0.001$ and confirming its universality across complex flow topologies.
\end{enumerate}

Beyond theoretical curiosity, this framework offers novel analytical pathways for complex physical challenges. The mathematical equivalence between diffusive entropy production and Fisher Information provides new optimization targets for chaotic advection in microfluidic mixing, and offers alternative, information-theoretic constraints for addressing the turbulence closure problem. To ensure the universality of this formulation, the analytical derivations are computationally validated against both one-dimensional models and a full three-dimensional, spatially varying Arnold--Beltrami--Childress (ABC) velocity field \cite{Dombre1986}. Furthermore, this framework yields analytical tools for assessing thermal entropy generation in heat exchangers, evaluating in-cylinder mixing in internal combustion engines, and predicting the physical thickness of aerodynamic shockwaves.

\section{Continuity Equation and Its Probabilistic Interpretation}
\label{sec:continuity}

Consider flow of a fluid of density $\rho$. Then, in a control volume, the general equation of continuity is given by
\begin{equation} \label{eq:continuity_general}
    \frac{\partial \rho}{\partial t} + \nabla \cdot (\rho \mathbf{v}) = 0,
\end{equation}
where $\rho$ represents the mass density of the fluid, $t$ represents time, $\mathbf{v}$ is the velocity vector field of the bulk fluid, and $\nabla \cdot$ denotes the divergence operator. In simple words, this equation represents the conservation of mass of fluid, mathematically. 

Now, the quantity $\rho(\mathbf{r},t)$ is the mass density. From the law of mass conservation, if I consider the entire control volume of the fluid flow, then,
\begin{equation} \label{eq:mass_integral}
    \int_{\mathrm{CV}} \rho(\mathbf{r},t)\,d\tau = M(t),
\end{equation}
where $M(t)$ is the total mass of the fluid that flows through the control volume $\mathrm{CV}$, and $d\tau$ is an infinitesimal volume element.

Now, given this relation, let's define,
\begin{equation} \label{eq:pdf_def}
    p(\mathbf{r};t) =
\begin{cases}
\dfrac{\rho(\mathbf{r},t)}{M(t)}, & \mathbf{r} \in \mathrm{CV}, \\
0, & \text{otherwise},
\end{cases}
\end{equation}
where $p(\mathbf{r};t)$ defines the spatial probability density function (PDF) parameterized by time $t$, and $\mathbf{r}$ is the spatial position vector. It can be shown that $p$ is the exact one-particle spatial probability density function of a fluid particle in a flow field \cite{HansenMcDonald2013}, and the proof is given in Appendix~\ref{app:proof}. 
Clearly,
\begin{equation} \label{eq:pdf_integral}
    \int_{\mathrm{\mathbb{R}^3}} p(\mathbf{r};t)\,d\tau = 1,
\end{equation}
where $\mathbb{R}^3$ represents the three-dimensional Euclidean space. This means that $p(\mathbf{r};t)$ is a valid probability density function. So, the continuity equation can be written as,
\begin{equation} \label{eq:continuity_prob}
    M(t)\left[\frac{\partial p}{\partial t} + \nabla \cdot (p\mathbf{v})\right] + \dot{M}(t)p = 0,
\end{equation}
where $\dot{M}(t)$ denotes the temporal rate of change (time derivative) of the total mass. The term $p\mathbf{v}$ may be thought as \textbf{probability flux} and $ \nabla \cdot (p\mathbf{v})$ is the \textbf{divergence of the flux}.

The underlying random variable may be viewed as the spatial location of a randomly sampled fluid particle at a given time \(t\). Equivalently, for an infinitesimal volume element \(d\tau\) centred at \(\mathbf{r}\), the random event is that the sampled particle is located within this volume at time \(t\). Thus, \(p(\mathbf{r};t)\,d\tau\) represents the probability that a randomly sampled particle occupies the volume \(d\tau\) around \(\mathbf{r}\) at time \(t\), establishing \(p(\mathbf{r};t)\) as the corresponding one-particle spatial probability density.

Therefore, the density of a fluid is modeled as a probability density function, and consequently, conservation of mass can be related to conservation of probability. This signifies that fluid motion can be viewed as transport of probability density in space and the velocity field governs how probability redistributes.

\section{Shannon Entropy and Its Evolution under Fluid Flow}
\label{sec:shannon_evolution}

\subsection{Definition of Shannon Entropy}
Consider a probability distribution with the pdf $p(\mathbf{r})$. For this distribution, the Shannon entropy is given by \cite{Shannon1948,CoverThomas2006}:
\begin{equation} \label{eq:shannon_def}
    H(t) = -\int_{\mathrm{\mathbb{R}^3}} p(\mathbf{r}) \,\log p(\mathbf{r}) \, d\tau,
\end{equation}
where $H(t)$ denotes the Shannon entropy of the system at time $t$, and $\log$ represents the natural logarithm. The Shannon entropy provides a quantitative measure of the uncertainty associated with a probability distribution. In the present context, it characterizes the spatial spread of the probability density $p(\mathbf{r};t)$.

\begin{remark}
Equation \eqref{eq:shannon_def} represents the continuous differential entropy of the probability density function $p(\mathbf r,t)$. Since $p$ carries physical dimensions of inverse volume, the logarithm implicitly requires normalization with respect to a reference density $p_0$ possessing identical dimensions. More rigorously, the entropy may be written as
\[
H(t)
=
-\int_{\mathbb R^3}
p(\mathbf r,t)
\log\left(
\frac{p(\mathbf r,t)}{p_0}
\right)d\tau.
\]
Using the normalization condition
\[
\int_{\mathbb R^3} p(\mathbf r,t)\,d\tau = 1,
\]
this differs from Equation \eqref{eq:shannon_def} only by the additive constant $\log p_0$. Consequently, all temporal entropy evolution equations derived in this work remain invariant under the choice of reference measure. In particular, the entropy production rate $\frac{dH}{dt}$ retains the well-defined physical dimensions $[T^{-1}]$.
\end{remark}

\subsection{Computation of Shannon Entropy in the Context of the Study}
\label{sec:computation_shannon}
I begin with differentiating $H$ with respect to time $t$:
\[
    \frac{dH}{dt}
= -\int_{\mathrm{\mathbb{R}^3}} \frac{\partial}{\partial t} \big( p \log p \big)\, d\tau,
\]
\begin{equation} \label{eq:dhdt_step1}
\text{or,}\quad\frac{dH}{dt}
= -\int_{\mathrm{\mathbb{R}^3}} (1 + \log p)\,\frac{\partial p}{\partial t}\, d\tau,
\end{equation}
where $\frac{dH}{dt}$ is the total temporal rate of change of the Shannon entropy.
Now substituting $\frac{\partial p}{\partial t} =- \nabla \cdot (p\mathbf{v})$, I obtain,
\begin{equation} \label{eq:dhdt_step2}
\frac{dH}{dt}
= \int_{\mathrm{\mathbb{R}^3}} (1 + \log p)\,\left[\nabla \cdot (p\mathbf{v})+\frac{\dot{M}}{M}p\right]\, d\tau.
\end{equation}
Now,
\[
    (1 + \log p)\,\nabla \cdot (p\mathbf{v})
    = \nabla \cdot \big[(1 + \log p)\,p\mathbf{v}\big]
    - p\mathbf{v} \cdot \nabla (1 + \log p).
\]
Therefore, I can rewrite,
\[
    \frac{dH}{dt}
    = \int_{\mathbb{R}^3} \nabla \cdot \big[(1 + \log p)\,p\mathbf{v}\big]\, d\tau
    - \int_{\mathbb{R}^3} p\mathbf{v} \cdot \nabla (1 + \log p)\, d\tau + \frac{\dot{M}}{M}\int_{\mathbb{R}^3} p(1+\log p) d\tau.
\]
And from Equations \eqref{eq:pdf_integral} and \eqref{eq:shannon_def}, this can be further rewritten as
\[
    \frac{dH}{dt}
    = \int_{\mathbb{R}^3} \nabla \cdot \big[(1 + \log p)\,p\mathbf{v}\big]\, d\tau
    - \int_{\mathbb{R}^3} p\mathbf{v} \cdot \nabla (1 + \log p)\, d\tau + \frac{\dot{M}}{M}(1-H).
\]

Noting that:
\[
\nabla (1 + \log p) = \frac{\nabla p}{p},
\]

Thus:
\[
\frac{dH}{dt}
= \int_{\mathbb{R}^3} \nabla \cdot \big[(1 + \log p)\,p\mathbf{v}\big]\, d\tau
- \int_{\mathbb{R}^3} \mathbf{v} \cdot \nabla p\, d\tau + \frac{\dot{M}}{M}(1-H),
\]
\[
\text{or,} \quad \frac{dH}{dt} = \int_{\mathbb{R}^3} \nabla \cdot  (p \mathbf{v}\,\log p)\, d\tau 
+ \int_{\mathbb{R}^3} p\,\nabla \cdot \mathbf{v}\, d\tau + \frac{\dot{M}}{M}(1-H).
\]
Applying the divergence theorem,
\[
\int_{\Omega} \nabla \cdot \mathbf{F}\, d\tau
= \int_{\partial \Omega} \mathbf{F}\cdot \mathbf{n}\, dS,
\]
where $\mathbf{F}$ is a continuously differentiable vector field, $\Omega$ is the volume domain, $\partial \Omega$ is the boundary surface of the volume, $\mathbf{n}$ is the outward-pointing unit normal vector, and $dS$ is the infinitesimal surface element, I obtain:
\[
\frac{dH}{dt}
= \int_{\partial \mathbb{R}^3} (p\,\mathbf{v}\,\log p)\cdot \mathbf{n}\, dS
+ \int_{\mathbb{R}^3} p\,\nabla \cdot \mathbf{v}\, d\tau +  \frac{\dot{M}}{M}(1-H).
\]
\begin{remark}
From the definition of $p$, i.e. from Equation \eqref{eq:pdf_def}, I may conclude that on the boundary of $\mathbb{R}^3$, $p=0$ as the control volume CV is finite and it has definitely no intersection with the boundary of $\mathbb{R}^3$.
\end{remark}
Hence,
\begin{equation} \label{eq:boundary_zero}
     \int_{\partial \mathbb{R}^3} (p\,\mathbf{v}\,\log p)\cdot \mathbf{n}\, dS = 0,
\end{equation}
considering that $p$ is continuous. Although in practice, this assumption may break down, giving rise to boundary terms in the right hand side of Equation \eqref{eq:boundary_zero}. The effect of the boundary terms are out of scope of the current study. 

Therefore, it may be concluded that,
\begin{equation} \label{eq:entropy_evolution_advective}
{{\frac{d}{dt}(MH)
= \dot{M} + M\int_{\mathrm{\mathbb{R}^3}} p \, \nabla \cdot \mathbf{v}\, d\tau}},
\end{equation}
where $MH$ represents the total mass-scaled spatial uncertainty.

At this juncture, it is necessary to rigorously define the term \textit{\textbf{steady flow}} as used throughout the remainder of this study. A flow is considered strictly steady if it simultaneously satisfies two conditions: the macroscopic velocity field is time-independent ($\frac{\partial \mathbf{v}}{\partial t} = 0$), and the net mass flux into the control volume is zero ($\dot{M} = 0$). So in this study, steady flow corresponds to a flow with a steady velocity field and and zero mass flux into the control volume. Under these steady conditions, Equation \eqref{eq:entropy_evolution_advective} reduces to: 

\begin{equation} \label{eq:entropy_evolution_advective_steady_state}
{{\frac{dH}{dt}
= \int_{\mathrm{\mathbb{R}^3}} p \, \nabla \cdot \mathbf{v}\, d\tau}}.
\end{equation}

The final result, i.e. Equation \eqref{eq:entropy_evolution_advective} is significant in the context of this particular study. This is because it represents that the \textbf{Shannon Entropy is dependent only on the divergence of the velocity field} (apart from the mass flow rate into the control volume and boundary terms) following the contribution of advective transport.  

\subsection{Shannon Entropy for Incompressible Flow}
For an incompressible flow, 
\begin{equation} \label{eq:incompressible_div}
    \nabla \cdot \mathbf{v} = 0.
\end{equation}
Therefore, for an incompressible flow,
\begin{equation} \label{eq:incompressible_entropy}
    \frac{d}{dt}(MH) = \dot{M},
\end{equation}
which signifies that, for an incompressible fluid flow, the Shannon entropy changes only due to the mass flow rate $\dot{M}$. Therefore, for a steady and incompressible flow, Shannon entropy remains constant with respect to time.

\subsection{Shannon Entropy for Compressible Flow}
The picture gets a little more complicated when the flow is not incompressible. In that case, $\nabla \cdot \mathbf{v} \neq 0$. The divergence of the velocity field can be both positive and negative. 

Now the equation of the continuity can be equivalently written as, 

\begin{equation} \label{eq:continuity_material}
     \frac{D\rho}{Dt} +  \rho \, \nabla \cdot\mathbf{v} = 0,
\end{equation}
where $\frac{D\rho}{Dt}$ represents the total (or material) derivative of the mass density.
This form directly reflects how the flow behaves with the sign of $\nabla \cdot \mathbf{v}$.

\subsubsection{Positive Divergence}
\label{sec:positive_divergence}

Let's look into the case of 
\begin{equation} \label{eq:pos_div}
    \nabla \cdot \mathbf{v} > 0.
\end{equation}

This means that,
\begin{equation} \label{eq:neg_Drho}
    \frac{D \rho}{Dt} < 0.
\end{equation}
Physically, this means that density of a fluid particle is decreasing, or in other words, \textbf{flow is expanding over time}. 

Now let's have a look at the Shannon entropy in this case. For $\nabla \cdot \mathbf{v}>0$, 
\begin{equation} \label{eq:pos_entropy_rate_mass}
    \frac {d}{dt}(MH)>\dot{M}.
\end{equation}

In an open system where total mass $M(t)$ fluctuates, the quantity $MH$ reflects the total mass-scaled spatial uncertainty of the fluid. This inequality indicates that the total spatial uncertainty of the system is growing at a rate that exceeds the baseline addition of new mass ($\dot{M}$). Physically, Shannon entropy represents the uncertainty in the outcome of a spatial random variable. The expanding flow field spreads the probability density function $p(\mathbf{r};t)$ over a larger domain, generating additional spatial uncertainty beyond what is simply carried into the system by the mass flux.

Also, it is worth noting that for a steady flow,
\begin{equation} \label{eq:pos_entropy_rate_steady}
    \frac {dH}{dt}>0,
\end{equation}
i.e. entropy increases with time \cite{Seifert2012,KondepudiPrigogine2014}.

\subsubsection{Negative Divergence}
This is the exact opposite case of what's discussed in sub-subsection \ref{sec:positive_divergence}, i.e. I have
\begin{equation} \label{eq:neg_div}
    \nabla \cdot \mathbf{v} < 0,
\end{equation}
and,
\begin{equation} \label{eq:pos_Drho}
    \frac{D \rho}{Dt} > 0.
\end{equation}
Physically, this means that density of a fluid particle is increasing, or in other words, the \textbf{flow is compressing over time}.

Similarly to the previous case, here,
\begin{equation} \label{eq:neg_entropy_rate_mass}
    \frac {d}{dt}(MH)<\dot{M}.
\end{equation}

This indicates that the rate of change of the total mass-scaled Shannon entropy is less than the rate of mass accumulation. Physically, a contracting flow field causes the probability density function $p(\mathbf{r};t)$ to become more localized to a specific region. This spatial localization reduces the spread of the distribution, which effectively lowers the overall spatial uncertainty relative to the mass entering the system.

Similar to the expanding flow, in this case, for a steady flow, I shall have,
\begin{equation} \label{eq:neg_entropy_rate_steady}
    \frac {dH}{dt}<0,
\end{equation}
i.e. entropy decreases with time.

\subsection{Interpretation of the Entropy Evolution Equation}

Therefore, I have established a direct connection between kinematics of fluid particle and entropy evolution. In particular, the divergence of the velocity field emerges as the governing quantity controlling entropy variation. Incompressible flows preserve entropy, indicating that the transport is purely redistributive and does not alter the overall uncertainty. In contrast, compressibility introduces local sources or sinks of entropy, leading to changes in the spatial spread of the probability density.

\section{Effect of Diffusion and Entropy Production}
\label{sec:diffusion}

Diffusive transport in fluid flow arises from microscopic random motion, leading to the net migration of particles from regions of higher concentration to lower concentration. This process tends to smooth out spatial gradients, resulting in a more uniform distribution over time. It is different from a uniform advective transport, and it has to be incorporated separately.

To distinguish the macroscopic continuum description from microscopic fluctuations, I write the probability density as \(p=p_{0}+\delta p\), where \(p_{0}\) denotes the smooth macroscopic component and \(\delta p\) represents localized fluctuations arising from the random motion of fluid molecules \cite{LandauLifshitz1969, BixonZwanzig1969}. At the macroscopic level, these fluctuations are neglected, since their ensemble average is assumed to vanish, \(\langle\delta p\rangle=0\); thus, the preceding analysis describes the \(p_{0}\) component alone. This distinction is very important because, in the mass conservation equation, there is no self diffusion of fluid particles in a single phase fluid. I now proceed to consider the microscopic diffusive transport associated with these fluctuations.

\subsection{Incorporation of Diffusion in Continuity Equation}

While the preceding analysis accounts for transport due to advection, most physical processes also involve diffusive effects arising from microscopic fluctuations. Thus far, the evolution of entropy has been examined under purely advective transport. I now extend the analysis to include diffusion, which introduces additional spreading of the probability density.

In broader terms, the conservation of probability at molecular level is given by,

\begin{equation} \label{eq:continuity_flux}
    M\left(\frac{\partial p}{\partial t} + \nabla \cdot \mathbf{J}\right)+\dot{M}p = 0,
\end{equation}
where $\mathbf{J}$ is defined as the total probability flux vector. It is critical to note that, in the mass conservation equation, there is no diffusive term for a single phase flow. However when I look at a molecular level, I need to incorporate for this diffusive transport, and therefore a diffusive term appears. Additionally, as mentioned before, $p$ is the one-particle spatial pdf, that governs the probability transport related to a single particle of the fluid.

When I incorporated only advective transport, I had, 
\[
    \mathbf{J} = p \mathbf{v}.
\]
Incorporating transport of particles due to diffusion, the flux becomes,
\begin{equation} \label{eq:flux_diffusion}
    \mathbf{J} = p \mathbf{v} - D \nabla p,
\end{equation}
where $D$ denotes the microscopic diffusion coefficient (diffusivity), and $\nabla p$ is the spatial gradient of the probability density. The diffusive flux $-D\nabla p$ is analogous to Fick's Law \cite{Risken1989}, and it represents the osmotic velocity of a fluid particle.

Therefore, now, the equation of continuity is given by,
\begin{equation} \label{eq:continuity_diffusion_full}
        M\left[\frac{\partial p}{\partial t} + \nabla \cdot (p\mathbf{v}-D\nabla p)\right]+\dot{M}p = 0.
\end{equation}
Assuming a homogeneous and isotropic medium with uniform properties, the diffusivity $D$ may be treated as a constant (with respect to space). With this assumption, the equation of continuity, incorporating diffusive transport, is,
\begin{equation} \label{eq:continuity_diffusion}
    {\frac{\partial p}{\partial t} + \nabla \cdot (p\mathbf{v}) +  \frac{\dot{M}}{M}p = D\,\nabla^2p},
\end{equation}
where $\nabla^2$ denotes the Laplacian operator.

\subsection{Computation of Shannon Entropy, Incorporating Diffusive Transport}

Substituting $\frac{\partial p}{\partial t} =- \nabla \cdot (p\mathbf{v})+D\,\nabla^2 p$ in Equation \eqref{eq:dhdt_step1}, I get,

\begin{equation} \label{eq:dhdt_diff_step1}
    \frac{dH}{dt}
= \int_{\mathrm{\mathbb{R}^3}} (1 + \log p)\,[\nabla \cdot (p\mathbf{v}) - D\,\nabla^2 p]\, d\tau +  \frac{\dot{M}}{M}(1-H),
\end{equation}
which can be written as
\begin{equation} \label{eq:dhdt_diff_step2}
    \frac{d}{dt}(MH)
= \dot{M} + M\left[\int_{\mathrm{\mathbb{R}^3}} (1 + \log p)\,\nabla \cdot (p\mathbf{v})\, d\tau\, - D\int_{\mathrm{\mathbb{R}^3}} (1 + \log p)\,\nabla^2 p\, d\tau\right].
\end{equation}

In Section \ref{sec:computation_shannon}, I have already shown that the first integral is equal to 
\[
    \int_{\mathbb{R}^3}p \, \nabla \cdot \mathbf{v} d\tau.
\]
So the second integral becomes our matter of interest in this section. Using the identity:
\[
\int_{\Omega} f\,\nabla^2 g\, d\tau
= -\int_{\Omega} \nabla f \cdot \nabla g\, d\tau
+ \int_{\partial \Omega} f\,\nabla g \cdot \mathbf{n}\, dS,
\]
where $f$ and $g$ are generic scalar functions defined over the domain $\Omega$, and with $f = 1 + \log p$ and $g = p$, I obtain:
\[
\int_{\mathbb{R}^3} (1+\log p)\,\nabla^2 p \, d\tau
= -\int_{\mathbb{R}^3} \nabla (1+\log p)\cdot \nabla p \, d\tau
+ \int_{\partial \mathbb{R}^3} (1+\log p)\,\nabla p \cdot \mathbf{n}\, dS.
\]

Now from Remark 2, the boundary integral vanishes (as $\nabla p$ decays faster than $\log p$). Therefore, only the first integral remains. Now, 
\[
    \int_{\mathbb{R}^3} \nabla (1+\log p)\cdot \nabla p \, d\tau = \int_{\mathbb{R}^3} \frac {\|\nabla p\|^2}{p}\, d\tau,
\]
where $\|\nabla p\|^2$ represents the squared Euclidean norm of the probability density gradient.

Hence, I get,
\begin{equation} \label{eq:entropy_evolution_diffusive}
    {\frac{d}{dt}(MH)= \dot{M} + M\left(\int_{\mathrm{\mathbb{R}^3}} p \, \nabla \cdot \mathbf{v}\, d\tau +  D\int_{\mathbb{R}^3} \frac {\|\nabla p\|^2}{p}\, d\tau\right)}.
\end{equation}

When the flow is steady, this equation reduces to:
\begin{equation} \label{eq:entropy_evolution_diffusive_steady_state}
    {\frac{dH}{dt}= \int_{\mathrm{\mathbb{R}^3}} p \, \nabla \cdot \mathbf{v}\, d\tau +  D\int_{\mathbb{R}^3} \frac {\|\nabla p\|^2}{p}\, d\tau}.
\end{equation}

\subsection{Interpretation of Diffusion as an Irreversible Process}

A key observation follows from the structure of the diffusive term.
\begin{remark}
    The integrand of the second integral is always positive, i.e.,
    \[
        \frac {\|\nabla p\|^2}{p}>0.
    \]
    Therefore, the integral (and hence the second term in the expression of $\frac{dH}{dt}$) is always positive, i.e.,
    \[
        D\int_{\mathbb{R}^3} \frac {\|\nabla p\|^2}{p}\, d\tau>0.
    \] 
    This means that, irrespective of the sign of the first integral, the rate of change of entropy w.r.t. time has a positive term. Due to that term, entropy is tending to increase with time.
\end{remark}

It is worth noting that the integral $\int_{\mathbb{R}^3} \frac{\|\nabla p\|^2}{p} d\tau$ corresponds to a form of \textbf{Fisher Information} of the spatial probability distribution \cite{Stam1959,Frieden2004}. Thus, Equation \eqref{eq:entropy_evolution_diffusive} reveals a profound physical symmetry: the rate of macroscopic entropy production due to fluid diffusion is directly proportional to the local Fisher Information of the flow field. This mirrors De Bruijn’s identity \cite{Stam1959} in pure information theory, now formally contextualized within fluid kinematics.

Now the second integral is solely due to the term due to diffusion in Equation \eqref{eq:continuity_diffusion}. Therefore, it may be said that diffusive transport in the medium causes a constant increase of entropy with time.

From the viewpoint of \textbf{thermodynamics}, a process in which the entropy increases is an \textbf{irreversible process}. Therefore, in fluid flow, \textbf{diffusive transport may be interpreted as an irreversible process}. 

It is important to distinguish the spatial Shannon entropy considered here from the Gibbs entropy defined over the full phase space of the system. The two quantities are not generally identical, since Gibbs entropy contains both configurational and momentum contributions. However, for an isothermal system with fixed particle masses and separable kinetic and configurational degrees of freedom, the momentum contribution remains constant, such that the temporal change in Gibbs entropy is determined by the change in the spatial entropy. Accordingly, $\frac{dS_{\mathrm{G}}}{dt} = k_{\mathrm{B}}\frac{dH}{dt},$ under these assumptions \cite{HnizdoGilson2010}. The connection is explored in details in Appendix~\ref{app:connection}.

\section{Physical Implications and Applications of the Result}
\label{sec:implications}

\subsection{Entropy Evolution in Canonical Flow Regimes}
Consider a steady and incompressible flow involving diffusive transport. In this case, the rate of Shannon entropy change is,
\begin{equation} \label{eq:entropy_canonical}
    \frac{dH}{dt}= D\int_{\mathbb{R}^3} \frac {\|\nabla p\|^2}{p}\, d\tau > 0.
\end{equation}
This expression is crucial for practices involving incompressible fluid flows. In theory, if I look at the macroscopic kinematics only, the fluid behaves like a perfectly reversible machine. If I could reverse the velocity field $\mathbf{v}$ to $-\mathbf{v}$, the macroscopic equations say the fluid should perfectly un-mix and go back to its starting position.

However, diffusion is inherently involved in almost all kinds of flows. Therefore, we'd still have a positive entropy generation rate from the diffusive transport. So, in terms of information science, \textbf{spatial information is being irreversibly degraded as the flow moves forward}. From the second law of thermodynamics, a process whose entropy generation is negative is not feasible. This is exactly why \textbf{we can't reverse a steady flow (without external energy) even if it's incompressible}.

\subsection{Compressibility, Spontaneity, and External Work}

\subsubsection{Rate of Entropy Change for Spontaneous Steady Flow}
From the second law of thermodynamics, the change of entropy in a spontaneous process is always non-negative. This directly complements the fact that \textbf{in isolated systems, fluids cannot compress spontaneously, as that will cause a decrease of entropy}.

From information theory, increasing Shannon entropy causes information loss. Therefore when $\nabla \cdot \mathbf{v}>0$, the kinematic term in the expression of $\frac {dH}{dt}$ acts as a certainty sink. Because the diffusion term is always positive, for spontaneous fluid flows,
\begin{equation} \label{eq:entropy_spontaneous}
    \frac{dH}{dt} > 0.
\end{equation}

Hence, it may be concluded that free expansion combined with diffusion is a spontaneous process. The system naturally evolves toward a state of maximum macroscopic and microscopic uncertainty.

\subsubsection{Requirement of External Work for Compressing Flow}

Now, compression of a fluid is an external-work driven process. So to get a negative value of $\frac{dH}{dt}$, I necessarily need external work, considering that the flow is steady. It is to be noted that in order to make $\frac{dH}{dt}$ negative, the external work has to overpower the term due to diffusion too, as that's always positive, as discussed earlier. 

From the reverse direction of this interpretation, I may conclude that, according to the Second Law of Thermodynamics, an isolated system cannot spontaneously decrease its entropy. Therefore, \textbf{achieving $dH/dt < 0$ in a steady fluid flow strictly requires the continuous input of external work} (e.g., a mechanical piston) to artificially force the fluid into a higher state of spatial certainty.

\subsubsection{Equilibrium Condition for Steady Flow}
\label{sec:equilibrium_condition}

For a steady flow, equilibrium can be defined as the case when entropy doesn't change with time, i.e. the flow is compressing in such a way that it completely balances the entropy generation due to diffusion. Therefore, the equilibrium condition is given by,
\begin{equation} \label{eq:equilibrium_integral}
    \int_{\mathrm{\mathbb{R}^3}} p \, \nabla \cdot \mathbf{v}\, d\tau =-  D\int_{\mathbb{R}^3} \frac {\|\nabla p\|^2}{p}\, d\tau.
\end{equation}
While Equation \eqref{eq:equilibrium_integral} guarantees that the net entropy of the entire system is unchanging, statistical mechanics allows for a more stringent condition \cite{Jaynes1957}. If I assume the principle of \textbf{Local Detailed Balance}—meaning the kinematic entropy sink perfectly cancels the diffusive entropy source at every infinitesimal point in the flow field—I can equate the integrands directly. So I have, 
\begin{equation} \label{eq:equilibrium_local}
    {\nabla \cdot \mathbf{v} =-  D\frac {\|\nabla p\|^2}{p^2}}.
\end{equation}

Equation \eqref{eq:equilibrium_local} is the mathematical representation of equilibrium condition in a flow involving diffusive transport. It is to be noted that Equation \eqref{eq:equilibrium_local} is a sufficient condition for a flow to be in equilibrium but it is not a necessary condition.

\begin{remark}
By defining a characteristic spatial thickness of the probability gradient as 
\begin{equation}
L_p = \frac{p}{||\nabla p||},
\end{equation}
where $L_p$ represents the characteristic length scale of the localized spatial uncertainty, Equation (34) simplifies to the explicit algebraic scaling law:
\begin{equation}
    |\nabla \cdot \mathbf{v}| = \frac{D}{L_p^2}.
\end{equation}

This relation reveals that the \textbf{flow reaches Local Detailed Balance precisely when the macroscopic compression time equates to the microscopic diffusion time, corresponding exactly to a local Péclet number of unity}. Consequently, this formulation provides a direct algebraic method to predict the physical thickness of compressive mixing layers and aerodynamic shockwaves purely from the local velocity divergence, bypassing the need for numerical integration.
\end{remark}

\subsection{Well Mixed State of Incompressible Flow}
Theoretically, the entropy for an incompressible flow will always keep on increasing. But if $H$ is still increasing as $t \to \infty$, that would mean that the probability is spreading out, which is not possible for a bounded flow. Therefore, I need $\nabla p=0$ at infinite time. 

Physically, $\nabla p= 0$ corresponds to the case of a \textbf{well mixed state}, i.e. I have a perfectly mixed, homogeneous fluid. As the Fisher Information term drops to zero, entropy generation stops.

\section{Computational Results and Discussion}
\label{sec:validation}

\subsection{Purely Diffusive Transport}
To verify the theoretical equivalence between the macroscopic entropy production rate and the local Fisher Information derived in Equation \eqref{eq:entropy_evolution_diffusive}, a one-dimensional numerical simulation of pure diffusive transport ($\mathbf{v} = 0$) was conducted. 

The one-dimensional continuous continuity equation, 
\begin{equation} \label{eq:fdm_diffusion}
    \frac{\partial p}{\partial t} = D \frac{\partial^2 p}{\partial x^2},
\end{equation}
where $x$ represents the one-dimensional spatial coordinate, was discretized using an explicit finite-difference method (FDM) with a forward-time, centered-space (FTCS) numerical scheme \cite{Tannehill1997}. The spatial domain was discretized with a uniform grid spacing of $\Delta x \approx 0.05$, and the temporal evolution was advanced using a time step of $\Delta t = 0.001$\,s, strictly satisfying the numerical stability criterion for the explicit scheme. The domain was initialized with a normalized Gaussian probability distribution to represent a localized fluid concentration. The boundaries of the finite numerical domain were strictly set to $p = 0$.

At each discrete time step, both sides of the entropy evolution equation were computed entirely independently:
\begin{enumerate}
    \item \textbf{The Left-Hand Side:} The total Shannon entropy $H(t)$ of the spatial grid was calculated, and its temporal rate of change $\dot H(t)$ was evaluated using numerical differentiation.
    \item \textbf{The Right-Hand Side:} The spatial density gradients $\nabla p$ were calculated using central differencing, and the Fisher Information integral $D \int_{\mathbb{R}^3} \frac{\|\nabla p\|^2}{p} d\tau$ was computed across the grid.
\end{enumerate}

\begin{figure}[h]
    \centering
    \includegraphics[width=0.9\textwidth]{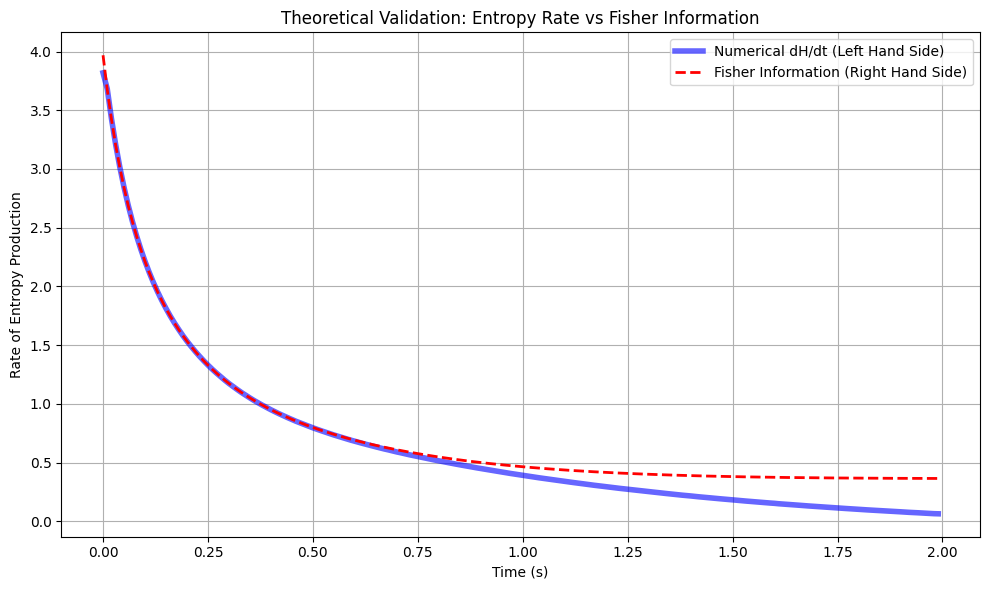}
    \caption{\small Comparison of the numerical rate of entropy production (Left-Hand Side) and the theoretical Fisher Information (Right-Hand Side) over time for a 1D diffusing Gaussian distribution. The precise overlap in the initial phase computationally validates the localized entropy source derived in Equation \eqref{eq:entropy_evolution_diffusive}. The divergence observed at $t > 0.6$\,s highlights the boundary effect as the probability distribution interacts with the finite computational domain, reinforcing the necessity of the unbounded spatial assumption.}
    \label{fig:pure_diffusion}
\end{figure}

As illustrated in Figure \ref{fig:pure_diffusion}, the independently computed numerical entropy rate and the theoretical Fisher Information curve exhibit perfect superimposition during the initial phase of the simulation. This agreement computationally validates the core theorem of this study: microscopic diffusion acts as a positive entropy source governed exactly by the local Fisher Information. 

Physically, the convex, decaying shape of the entropy production curve is driven by the spatial flattening of the probability distribution. As the localized fluid concentration diffuses outward, the spatial density gradients ($\nabla p$) become progressively shallower. Because the diffusive entropy source is proportional to the square of these gradients ($||\nabla p||^2$), the rate of information loss naturally diminishes as the fluid physically approaches a more homogeneous, well-mixed state.

A deliberate deviation is observed at later time steps. This divergence occurs precisely when the tails of the diffusing probability distribution begin to interact with the finite boundaries of the computational grid. This physical collision violates the unbounded space ($\mathbb{R}^3$) assumption established in Remark 2, causing probability to artificially exit the control volume. Rather than a flaw in the model, this boundary effect computationally reinforces the necessity of the boundary-vanishing assumption for perfect information preservation.

\subsection{Transport Involving non-zero Velocity Field}
To observe the interplay between the kinematic entropy sink and the diffusive entropy source, a steady compressive velocity field was defined as $v(x) = -cx$, where $c > 0$ represents a constant strain rate. This yields a steady, spatially uniform negative divergence, $\nabla \cdot \mathbf{v} = -c$. The corresponding one-dimensional continuity equation is given by:
\begin{equation} \label{eq:fdm_advection_diffusion}
\frac{\partial p}{\partial t} + \frac{\partial}{\partial x}(p v) = D \frac{\partial^2 p}{\partial x^2}.
\end{equation}
This continuous partial differential equation was discretized using an explicit finite-difference method (FDM). To maintain numerical stability in the presence of advection, an upwind differencing scheme was applied to the convective term, while a centered-space scheme was retained for the diffusive term. The spatial domain was discretized with a uniform grid spacing of $\Delta x = 0.01$, and the temporal evolution was advanced using a time step of $\Delta t = 0.0001$\,s, strictly satisfying the stability criteria for the selected scheme. The spatial domain was initialized with a normalized Gaussian probability distribution, representing a localized fluid concentration. The computational domain was chosen to be sufficiently large such that the probability density at the boundaries remained strictly $p = 0$ throughout the simulation duration, ensuring that the net mass flux into the domain was zero ($\dot{M} = 0$). 

At each discrete time step, both sides of the steady-state entropy evolution equation were computed independently across the spatial grid:
\begin{enumerate}
\item \textbf{The Left-Hand Side:} The total Shannon entropy $H(t)$ of the spatial grid was calculated, and its temporal rate of change $\frac{dH}{dt}$ was evaluated using forward numerical differentiation.
\item \textbf{The Right-Hand Side:} The spatial density gradients $\nabla p$ were calculated using central differencing. The total entropy generation rate was computed as the sum of the constant kinematic sink integral $-c \int_{\mathbb{R}^3} p \, d\tau$ (which evaluates to $-c$) and the diffusive Fisher Information integral $D \int_{\mathbb{R}^3} \frac{\|\nabla p\|^2}{p} d\tau$.
\end{enumerate}

\begin{figure}[h]
\centering
\includegraphics[width=0.9\textwidth]{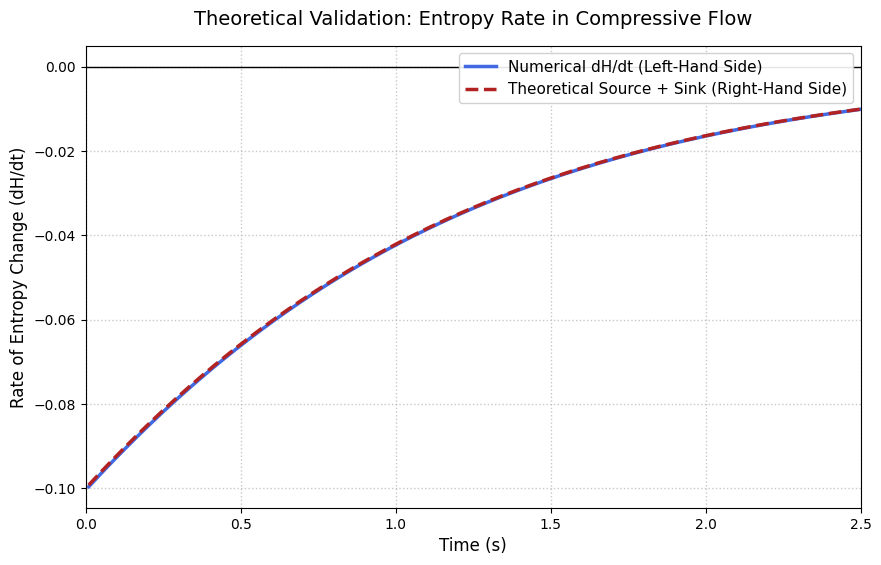}
\caption{\small Comparison of the numerically evaluated $\frac{dH}{dt}$ (Left-Hand Side) and the analytically constructed summation of the kinematic sink and Fisher Information source (Right-Hand Side) for a 1D compressive flow. The initial negative phase demonstrates the spatial dominance of the compressive kinematic sink, while the asymptotic approach toward $dH/dt = 0$ computationally validates the Local Detailed Balance equilibrium condition derived in Equation \eqref{eq:equilibrium_local}.}
\label{fig:compressive_flow}
\end{figure}

As observed in Figure \ref{fig:compressive_flow}, the net rate of entropy change ($\frac{dH}{dt}$) is initially negative. This indicates that the constant compressive kinematic sink spatially dominates the initial diffusive entropy source, resulting in a net localization of the probability density. Physically, the macroscopic compression acts as a kinematic funnel, continuously packing the fluid probability into a contracting region. This artificial spatial confinement forcibly steepens the density gradients ($\nabla p$), which in turn exponentially amplifies the microscopic diffusive pushback. 

The consistent overlap of the independently computed numerical and theoretical curves validates Equation \eqref{eq:entropy_evolution_diffusive_steady_state}. Furthermore, the asymptotic convergence toward $\frac{dH}{dt} = 0$ illustrates the system naturally approaching the equilibrium condition of Local Detailed Balance. Physically, this plateau represents the exact moment the outward `pressure' of microscopic diffusion perfectly balances the inward macroscopic crush of advection, causing the fluid to dynamically lock into a steady state. The fact that the two curves overlap completely further goes on to support the local detailed balance derived in Section \ref{sec:equilibrium_condition}.

\subsection{Three-Dimensional Transport in a Spatially Varying Velocity Field}
To verify the robustness of Equation (30) beyond one-dimensional canonical flows, a full three-dimensional computational validation was performed. The advective-diffusive transport was simulated under a highly non-trivial, spatially varying velocity field.

The field was constructed by superimposing a three-dimensional compressive perturbation onto an incompressible Arnold–Beltrami–Childress (ABC) type flow. The velocity components are given by:

\begin{equation}
v_x = v_0 \sin\left(\frac{2\pi y}{L}\right) \cos\left(\frac{2\pi z}{L}\right) + \varepsilon v_0 \sin\left(\frac{2\pi x}{L}\right),
\end{equation}
\begin{equation}
v_y = v_0 \sin\left(\frac{2\pi z}{L}\right) \cos\left(\frac{2\pi x}{L}\right) + \varepsilon v_0 \sin\left(\frac{2\pi y}{L}\right),
\end{equation}
\begin{equation}
v_z = v_0 \sin\left(\frac{2\pi x}{L}\right) \cos\left(\frac{2\pi y}{L}\right) - 2\varepsilon v_0 \sin\left(\frac{2\pi z}{L}\right),
\end{equation}
where $v_x$, $v_y$, and $v_z$ represent the Cartesian components of the velocity field, $v_0$ is the base velocity amplitude, $L$ is the characteristic domain length scale, and $\varepsilon$ is the dimensionless amplitude of the compressive perturbation mode. This configuration ensures that the divergence of the velocity field, $\nabla \cdot \mathbf{v}$, is non-zero and varies continuously across the spatial domain, activating both the kinematic sink/source term and the Fisher Information source term simultaneously.

The continuous partial differential equation was discretized using a fully conservative central-difference finite-difference scheme. The spatial domain was discretized with uniform grid spacings of $\Delta x = \Delta y = \Delta z \approx 0.0833$, and the temporal evolution was advanced using a time step of $\Delta t = 0.001$\,s, strictly satisfying the necessary numerical stability conditions. This approach strictly preserves the spatial information without introducing artificial numerical diffusion (artificial viscosity), ensuring a mathematically rigorous evaluation of the theoretical entropy evolution.

\begin{figure}[htbp]
    \makebox[\textwidth][c]{%
        \includegraphics[width=1\textwidth,height=0.78\textheight]{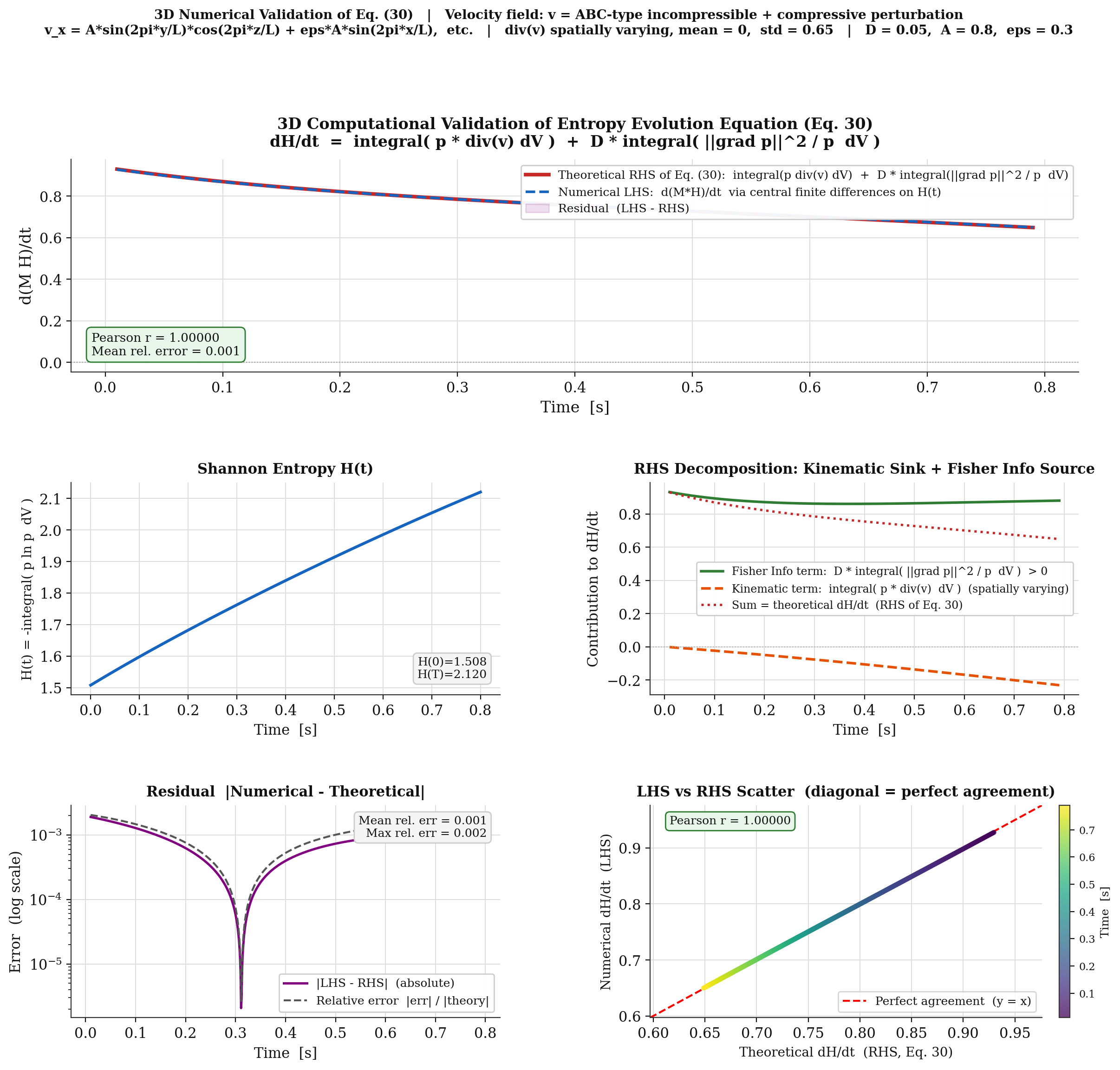}%
    }
    \caption{Computational validation of Equation (30) under a 3D spatially varying ABC-type velocity field. The near-perfect superimposition (mean relative error = 0.001) of the numerical entropy derivative and the theoretically computed kinematic and Fisher Information sources validates the universality of the derived framework.}
    \label{fig:3d_validation}
\end{figure}

At each discrete time step, the temporal rate of change of the mass-scaled spatial Shannon entropy (Left-Hand Side) was computed independently from the volume integration of the kinematic and diffusive sources (Right-Hand Side). As illustrated in Figure \ref{fig:3d_validation}, the numerical validation is presented across five distinct sub-analyses, each capturing a different physical facet of the fluid's informational transport:

\begin{itemize}
    \item \textbf{Global Entropy Evolution (Top Panel):} The numerically computed rate of entropy production perfectly superimposes with the analytically predicted summation of the kinematic and Fisher Information terms. Physically, this confirms that macroscopic chaotic advection and microscopic diffusive spreading collectively dictate the global information loss, with no "missing" thermodynamic mechanics.
    
    \item \textbf{Absolute Shannon Entropy $H(t)$ (Middle Left):} The monotonic increase of the absolute spatial entropy from $H(0)=1.508$ to $H(T)=2.120$ physically represents the irreversible degradation of spatial certainty. The chaotic streamlines of the ABC flow continuously stretch and fold the fluid elements, permanently increasing the global spatial uncertainty.
    
    \item \textbf{RHS Decomposition (Middle Right):} This panel captures the core physical ``tug-of-war" within the flow. The spatially varying compressive perturbation acts as a localized \textit{kinematic sink} (the negative, orange dashed line), artificially attempting to pack the fluid and decrease entropy. However, this localized packing violently steepens the spatial density gradients ($\nabla p$). Because the diffusive source is proportional to $||\nabla p||^2$, the \textit{Fisher Information} (the positive, solid green line) exponentially amplifies in response. The physics strictly obey the Second Law of Thermodynamics: the positive diffusive pushback perpetually overpowers the kinematic compression, ensuring a net positive entropy generation rate.
    
    \item \textbf{Residual Analysis (Bottom Left):} The absolute and relative errors remain exceptionally low (mean relative error of $0.001$), confirming that the entropy generation observed is purely a physical consequence of the derived kinematics, rather than an artifact of artificial numerical diffusion. 
    
    \item \textbf{LHS vs RHS Scatter (Bottom Right):} The perfect linear correlation (Pearson $r = 1.00000$) lying strictly along the diagonal establishes that the theoretical framework holds uniformly across time. Regardless of the instantaneous complexity, stretching, or localized compression of the 3D velocity field, the localized informational mechanics dictate the exact thermodynamic trajectory of the fluid.
\end{itemize}

Ultimately, this computational result provides robust, generalized validation that the localized entropy generation mechanisms derived in this study hold universally for complex, three-dimensional flow topologies. The framework successfully captures the fundamental thermodynamic reality of continuous media: macroscopic kinematic features (like compressive sinks and chaotic advective stretching) perpetually manufacture steep gradient interfaces, which microscopic diffusion instantly acts upon to irreversibly drive macroscopic entropy production.

\section{Proposed Applications and Future Directions}
\label{applications}
Thus far, I have presented a hybrid model of a fluid flow using probability and information science. It is important to explicitly state that the following subsections present theoretical extrapolations and proposed future research directions rather than computationally or experimentally validated results. There are several complex engineering systems where this informational framework offers novel analytical constraints, which are proposed here as foundational concepts to be tested and validated in future studies.

\subsection{Implications for Turbulence Modeling}
In turbulent flows, energy cascades from large eddies down to microscopic scales where it is destroyed by viscosity. Modeling this `closure problem' in the Navier-Stokes equations is notoriously difficult \cite{Peters2000} and usually relies on heavy empirical tuning. However, from another angle, turbulence can be viewed as an intense mixing process that maximizes entropy generation. This is because turbulent flows are more random than any possible laminar flow. So if $D_{t}$ is the turbulent eddy diffusivity, and assuming a stationary macroscopic domain where net mass flux $\dot{M}=0$, then the velocity field $\mathbf{v}^{*}$ is such that:

\begin{equation}
\mathbf{v}^{*} = \arg\max_{\mathbf{v}} \dot{H}(t) = \arg\max_{\mathbf{v}} \left[ \int_{\mathbb{R}^{3}} p\nabla\cdot \mathbf{v}~d\tau + D_{t}\int_{\mathbb{R}^{3}} \frac{||\nabla p||^{2}}{p} d\tau \right],
\end{equation}
where $\mathbf{v}^{*}$ denotes the theoretical optimal velocity field that maximizes the rate of entropy production, $D_{t}$ represents the turbulent eddy diffusivity, and $\arg\max$ indicates the argument that maximizes the given objective function. Now, actually for high $\text{Re}$ flows, the term due to diffusion in the expression of $H(t)$ is much greater than the term due to advection. Therefore, incorporating this fact, from the perspective of this study, I am looking for the velocity field $\mathbf{v}^{*}$ that maximizes the gradient of the pdf $p$, i.e. $\nabla p$. This may cause confusion, as macroscopically, advection dominates diffusion in high $\text{Re}$ flows. At high Reynolds numbers, the intense advection creates an energy cascade. The large eddies break down into smaller and smaller eddies, eventually reaching microscopic sizes called the Kolmogorov microscales.

At these microscopic scales, the fluid is stretched so violently that the physical distance between high and low density approaches zero. Therefore, the gradient $||\nabla p||^{2}$ explodes toward infinity. Because the gradient is squared in the Fisher Information term, this term dominates the entropy production rate. Hence, this informational model offers a complementary theoretical constraint that could inform future turbulence closure schemes \cite{MajdaKleeman2002}.

\subsection{Heat Exchanger Design and Thermal Entropy Generation}
In thermal management systems such as heat exchangers, the primary objective is to maximize heat transfer, which inherently relies on effective fluid mixing and the creation of steep thermal gradients. Traditional thermal design often utilizes Entropy Generation Minimization (EGM) to balance heat transfer efficiency against viscous pressure losses \cite{Bejan1996}. By mapping localized fluid temperature variations to spatial probability variations, the framework presented in this study offers an alternative analytical perspective. Let $\theta(\mathbf{r})$ be the non-dimensionalized temperature field of an incompressible flow. A thermal spatial probability density function, $p_{T}$, may be defined as:

\begin{equation}
p_{T}(\mathbf{r})=\frac{\theta(\mathbf{r})}{\int_{\Omega}\theta(\mathbf{r})d\tau},
\end{equation}
where $p_T(\mathbf{r})$ is the thermal spatial probability density function, $\theta(\mathbf{r})$ denotes the non-dimensionalized localized temperature field, and $\Omega$ is the internal volume of the heat exchanger. For an incompressible flow ($\nabla\cdot \mathbf{v}=0$), which is generally the case in most heat exchangers, the kinematic sink term vanishes. Substituting the molecular diffusivity with the thermal diffusivity $\alpha$, the spatial thermal entropy evolution is governed entirely by the Fisher Information term:

\begin{equation}
\frac{dH_{T}}{dt}=\alpha\int_{\Omega}\frac{||\nabla p_{T}||^{2}}{p_{T}}d\tau,
\end{equation}
where $H_T$ represents the thermal spatial Shannon entropy, and $\alpha$ is the thermal diffusivity coefficient. This relation serves as a direct mathematical quantifier of spatial mixing intensity. Instead of relying solely on empirical heat transfer coefficients, engineers could theoretically optimize the placement of internal geometries (such as baffles and fins) by treating the Fisher Information integral as an objective function. Maximizing this term in specific geometric regions corresponds to maximizing the local mixing rate, providing a purely kinematic and informational method for evaluating and optimizing heat exchanger efficiency.

\subsection{Internal Combustion Engine In-Cylinder Dynamics}
Internal combustion engines present a highly transient fluid environment characterized by both macroscopic compressibility and intense microscopic turbulent mixing. During the compression stroke, the upward motion of the piston enforces a negative divergence ($\nabla\cdot \mathbf{v}<0$) throughout the cylinder. According to the derived framework, this macroscopic kinematic compression acts as an informational sink, localizing the air-fuel probability density into a decreasing volume. Simultaneously, the high-pressure turbulent injection of fuel creates sharp spatial concentration gradients ($\nabla p$) which act as a diffusive entropy source. To quantify this analytically, let $p_{f}(\mathbf{r},t)$ represent the normalized spatial probability density of the fuel mass fraction within the instantaneous cylinder volume $V(t)$. Assuming the macroscopic continuity dictates a uniform negative divergence proportional to the volumetric strain rate, $\nabla\cdot \mathbf{v}\approx\frac{\dot{V}(t)}{V(t)}$, and substituting the molecular diffusivity with a turbulent eddy diffusivity $D_{t}$, the evolution of the fuel's spatial uncertainty may be written as:

\begin{equation}
\frac{dH_{f}}{dt}=\frac{\dot{V}(t)}{V(t)}+D_{t}\int_{V(t)}\frac{||\nabla p_{f}||^{2}}{p_{f}}d\tau,
\end{equation}
where $H_f$ is the fuel spatial uncertainty (entropy), $V(t)$ is the instantaneous cylinder volume, $\dot{V}(t)$ is the volumetric strain rate, and $D_t$ represents the turbulent eddy diffusivity. In this expression, the first term represents the kinematic sink (which is negative due to $\dot{V}<0$), while the second term represents the positive Fisher Information source associated with turbulent fuel dispersal. An effective ignition event generally requires a highly mixed, uniform charge. As turbulent mixing progresses, concentration gradients flatten, causing the Fisher Information term to decay. The theoretical equilibrium state of the mixture where turbulent homogenization is balanced by macroscopic compression occurs when $\frac{dH_{f}}{dt}=0$. This balance yields an explicit mathematical criterion that could potentially be used to estimate optimal spark-ignition timing ($t_{\text{spark}}$):

\begin{equation}
\frac{\dot{V}(t_{\text{spark}})}{V(t_{\text{spark}})}=-D_{t}\int_{V(t_{\text{spark}})}\frac{||\nabla p_{f}||^{2}}{p_{f}}d\tau,
\end{equation}
where $t_{\text{spark}}$ signifies the theoretical optimal spark-ignition timing. This condition provides a localized, information-theoretic target to define the mixed state of the fuel charge. By linking the macroscopic engine kinematics directly to the scalar dissipation rate of the fuel, this formulation may offer engine designers an alternative analytical method to supplement computationally intensive three-dimensional Navier-Stokes simulations.

\subsection{Optimization of Microfluidic Mixing}
When dealing with microscopic fluid flows (like blood in a micro-capillary or chemicals in a diagnostic chip), the flow is strictly laminar ($\text{Re}\ll1$). Mixing two fluids together is incredibly difficult because there is no turbulence to help; I have to rely purely on advective stretching and molecular diffusion. Fluids like water or blood are incompressible in practice. So for these microfluidic flows, Equation (31) governs mixing. It is important to note that, a higher rate of entropy production is generally associated with reduced mixing times. I cannot change $D$; but $||\nabla p||$ can be maximized, by optimizing the physical geometry of the micro-channels to induce chaotic advection \cite{Aref1984}. Thus, laminar flows can be engineered in such a way that the mixing time is minimized.

\subsection{Shockwave Profiling in High Speed Aerodynamics}
In classical aerodynamics, when a fluid moves at supersonic speeds and suddenly hits an obstacle, it forms a shockwave. Macroscopically, aerodynamic engineers treat a shockwave as a perfect, infinitely thin discontinuity. The fluid instantly jumps from low density to high density, and from high velocity to low velocity. But in reality, nature doesn't allow perfect discontinuities. A shockwave actually has a measurable physical thickness (usually on the order of a few mean free paths of the gas molecules). The extreme compression of the flow ($\nabla\cdot \mathbf{v}\ll0$) is desperately trying to smash the shockwave into an infinitely thin line. However, this creates a massive density gradient ($\nabla p$), which causes thermal and viscous diffusion ($D$) to aggressively push back and smear the wave out. Instead of relying purely on expensive supersonic wind tunnel experiments or heavy numerical simulations to figure out what happens inside a shockwave, Equation (34) may be used as a basis for analytical estimation. By plugging in the known velocity field ($\mathbf{v}$) and the fluid's diffusivity ($D$), I can integrate this differential equation to estimate the theoretical density profile $p(\mathbf{r})$ and the precise physical thickness of the shock layer.

\section{Limitations of the Study}
While the presented informational framework provides novel thermodynamic insights into macroscopic fluid transport, it is subject to several fundamental limitations that constrain its immediate applicability to certain complex systems:

\begin{itemize}
    \item \textbf{Constant Scalar Diffusivity:} The derivation assumes a homogeneous, isotropic fluid medium characterized by a constant scalar diffusivity, $D$. In highly anisotropic flows or turbulent regimes with directionally dependent mixing scales, the diffusivity must be modeled as a second-order tensor, $\mathbf{D}$, which would complicate the scalar Fisher Information integral.
    
    \item \textbf{Unbounded Spatial Domain:} The mathematical guarantee of a strictly positive entropy source relies heavily on the unbounded spatial domain assumption (Remark 2), where probability gradients decay to zero at infinity. As explicitly demonstrated by the boundary collision effect in the numerical simulation (Section \ref{sec:validation}.1), applying this framework to strictly bounded domains with finite physical walls requires the formulation of additional boundary flux terms to prevent artificial information leakage.
    
    \item \textbf{Restriction to Fickian Diffusion:} The incorporation of microscopic diffusion in this study strictly follows Fick's Law. Consequently, this model cannot currently accurately capture anomalous or non-Fickian transport phenomena (such as sub-diffusion or super-diffusion), which are frequently encountered in complex, heterogeneous porous media or viscoelastic fluids.
    
    \item \textbf{Purely Kinematic Formulation:} The current model is purely kinematic. It successfully evaluates the entropy production of a \textit{given} velocity field $\mathbf{v}$, but it does not inherently couple back to the Navier-Stokes momentum equations. To fully realize this framework as a predictive fluid dynamics tool, future formulations must explicitly couple the spatial probability gradients to physical pressure gradients and viscous stress tensors.
\end{itemize}

\section{Conclusion}
This paper has established a formal mathematical framework that bridges classical fluid mechanics with information theory by interpreting macroscopic fluid density as a spatial probability density function. Through this interdisciplinary lens, fluid dynamics is mathematically redefined as the continuous transport of probability and evolution of uncertainty.

By deriving the exact temporal evolution of Shannon entropy under general flow conditions, I have demonstrated that macroscopic advection acts as an information-preserving transport mechanism, while flow compressibility ($\nabla\cdot \mathbf{v}$) serves as a kinematic source or sink of spatial certainty. Most crucially, the incorporation of microscopic diffusion mathematically guarantees a strictly positive rate of entropy production, as discussed in Remark 3. I have shown that this irreversible loss of information is governed directly by the local Fisher Information, securely anchoring the physical ``arrow of time'' within the fundamental kinematics of fluid flow.

This informational perspective of fluid flow yields a rigorous thermodynamic classification of fluid processes. Spontaneous fluid expansion unconditionally maximizes spatial entropy, whereas non-spontaneous compression mandates the input of external work to artificially overcome the diffusive destruction of information. Furthermore, by defining the state of constant spatial entropy, I derived the explicit condition for Local Detailed Balance, which elegantly reduces to an algebraic scaling law for predicting the thickness of aerodynamic shockwaves and compressive layers at a local P\'{e}clet number of unity. To guarantee the robustness of these analytical findings, the framework was computationally validated. The near-perfect superimposition of the numerical entropy derivative and the theoretical Fisher Information source under a fully three-dimensional, compressive ABC-type flow confirms the universality of the derived equations across complex flow topologies.

Apart from theoretical unification, this framework presents novel analytical pathways for fluid phenomena, as discussed in Section \ref{applications}. The mathematical equivalence between diffusive entropy production and Fisher Information offers fundamental constraints for turbulence modeling and provides new objective functions for optimizing scalar mixing in microfluidics. Ultimately, viewing Fluid Mechanics through the paradigm of Information Science not only deepens our fundamental understanding of thermodynamic irreversibility, but also equips researchers with powerful new thermodynamic tools to model and predict continuous media phenomena.

\section*{Data and Code Availability}
The C++ source code utilized for the finite-difference numerical simulations (modeling both the pure diffusion and the steady compressive flow), the Python script used for data visualization, and the generated numerical datasets supporting the computational validation of this study are openly available in a public GitHub repository at: \url{https://github.com/SrabonGitikar/Entropy-Diffusion-FDM}.

\section*{Acknowledgements}
I gratefully acknowledge Katha Ganguly (IISER Pune) for her careful review of the manuscript prior to submission and for her valuable feedback. I also thank the reviewer(s) for their constructive comments and insightful suggestions, which significantly improved the quality of this study.

\bibliographystyle{unsrt}
\bibliography{references}

@book{Som2012,
  author    = {Som, S. K. and Biswas, G. and Chakraborty, S.},
  title     = {Introduction to Fluid Mechanics and Fluid Machines},
  publisher = {McGraw-Hill Education},
  year      = {2012}
}

@book{White2011,
  author    = {White, F. M.},
  title     = {Fluid Mechanics},
  publisher = {McGraw-Hill},
  year      = {2011}
}

@article{IcardiDentz2019,
  author  = {Icardi, M. and Dentz, M.},
  title   = {Probability density function (PDF) models for particle transport in porous media},
  journal = {Physical Review Fluids},
  volume  = {4},
  year    = {2019},
  pages   = {124501}
}

@article{Pope1985,
  author  = {Pope, S. B.},
  title   = {PDF methods for turbulent reactive flows},
  journal = {Progress in Energy and Combustion Science},
  volume  = {11},
  year    = {1985},
  pages   = {119--192}
}

@article{Boujelbene2024,
  author = {Boujelbene, M. and Rehman, S. and Hashim and Balegh, M.},
  title   = {Entropy degradation in a dual diffusion flow of a non-Newtonian fluid in inclined channel using Keller-Box approach},
  journal = {Mathematical Methods in the Applied Sciences},
  volume  = {48},
  year    = {2024}
}

@article{Alqahtani2023,
  author  = {Alqahtani, S. and Rehman, S. and Alshehery, S.},
  title   = {Computational method for energy transport of MHD nanofluids flow near non-aligned stagnation point with non-linear thermal radiation and interface slip},
  journal = {Results in Engineering},
  volume  = {19},
  year    = {2023},
  pages   = {101383}
}

@article{Khedher2023,
  author  = {Khedher, N. B. and Rehman, S. and Alqahtani, S. and Alshehery, S.},
  title   = {Comparative study of entropy distribution for generalized fluid between an inclined channel in the perspective of classical and non-Fourier's law},
  journal = {Engineering Science and Technology, an International Journal},
  volume  = {45},
  year    = {2023},
  pages   = {101471}
}

@article{Dombre1986,
  author  = {Dombre, T. and Frisch, U. and Greene, J. M. and H{\'e}non, M. and Mehr, A. and Soward, A. M.},
  title   = {Chaotic streamlines in the ABC flows},
  journal = {Journal of Fluid Mechanics},
  volume  = {167},
  year    = {1986},
  pages   = {353--391}
}

@book{HansenMcDonald2013,
  author    = {Hansen, Jean-Pierre and McDonald, Ian R.},
  title     = {Theory of Simple Liquids: With Applications to Soft Matter},
  edition   = {4},
  publisher = {Academic Press},
  year      = {2013},
  address   = {Oxford},
  doi       = {10.1016/C2010-0-66723-X}
}

@article{Shannon1948,
  author  = {Shannon, C. E.},
  title   = {A mathematical theory of communication},
  journal = {The Bell System Technical Journal},
  volume  = {27},
  year    = {1948},
  pages   = {379--423}
}

@book{CoverThomas2006,
  author    = {Cover, T. M. and Thomas, J. A.},
  title     = {Elements of Information Theory},
  publisher = {John Wiley \& Sons},
  year      = {2006}
}

@book{LandauLifshitz1969,
  author    = {Landau, L. D. and Lifshitz, E. M.},
  title     = {Statistical Physics, Part 1},
  edition   = {2},
  publisher = {Pergamon Press},
  address   = {Oxford},
  year      = {1969}
}

@article{BixonZwanzig1969,
  author  = {Bixon, M. and Zwanzig, R.},
  title   = {Boltzmann-Langevin equation and hydrodynamic fluctuations},
  journal = {Physical Review},
  volume  = {187},
  number  = {1},
  year    = {1969},
  pages   = {267--272},
  doi     = {10.1103/PhysRev.187.267}
}

@book{Risken1989,
  author    = {Risken, H.},
  title     = {The Fokker--Planck Equation: Methods of Solution and Applications},
  publisher = {Springer},
  year      = {1989}
}

@article{Stam1959,
  author  = {Stam, A. J.},
  title   = {Some inequalities satisfied by the quantities of information of Fisher and Shannon},
  journal = {Information and Control},
  volume  = {2},
  year    = {1959},
  pages   = {101--112}
}

@book{Frieden2004,
  author    = {Frieden, B. R.},
  title     = {Science from Fisher Information: A Unification},
  publisher = {Cambridge University Press},
  year      = {2004}
}

@article{Jaynes1957,
  author  = {Jaynes, E. T.},
  title   = {Information theory and statistical mechanics},
  journal = {Physical Review},
  volume  = {106},
  year    = {1957},
  pages   = {620}
}

@article{HnizdoGilson2010,
  author  = {Hnizdo, Vladimir and Gilson, Michael K.},
  title   = {Thermodynamic and Differential Entropy Under a Change of Variables},
  journal = {Entropy},
  volume  = {12},
  number  = {3},
  pages   = {578--590},
  year    = {2010},
  doi     = {10.3390/e12030578}
}

@article{Seifert2012,
  author  = {Seifert, U.},
  title   = {Stochastic thermodynamics, fluctuation theorems and molecular machines},
  journal = {Reports on Progress in Physics},
  volume  = {75},
  year    = {2012},
  pages   = {126001}
}

@book{KondepudiPrigogine2014,
  author    = {Kondepudi, D. and Prigogine, I.},
  title     = {Modern Thermodynamics: From Heat Engines to Dissipative Structures},
  publisher = {John Wiley \& Sons},
  year      = {2014}
}

@book{Tannehill1997,
  author    = {Tannehill, J. C. and Anderson, D. A. and Pletcher, R. H.},
  title     = {Computational Fluid Mechanics and Heat Transfer},
  publisher = {Taylor \& Francis},
  year      = {1997}
}

@book{Peters2000,
  author    = {Peters, N.},
  title     = {Turbulent Combustion},
  publisher = {Cambridge University Press},
  year      = {2000}
}

@article{MajdaKleeman2002,
  author  = {Majda, A. J. and Kleeman, R.},
  title   = {Measuring information content in model error and prediction},
  journal = {Journal of the Atmospheric Sciences},
  volume  = {59},
  year    = {2002},
  pages   = {1257--1274}
}

@book{Bejan1996,
  author    = {Bejan, A.},
  title     = {Entropy Generation Minimization: The Method of Thermodynamic Optimization of Finite-Size Systems and Finite-Time Processes},
  publisher = {CRC Press},
  year      = {1996}
}

@article{Aref1984,
  author  = {Aref, H.},
  title   = {Stirring by chaotic advection},
  journal = {Journal of Fluid Mechanics},
  volume  = {143},
  year    = {1984},
  pages   = {1--21}
}

\appendix

\section{Proof that $p$ is the One-Particle Spatial Probability Density}
\label{app:proof}

Consider a system of \(N\) fluid particles with positions
\(\mathbf{r}_1(t),\ldots,\mathbf{r}_N(t)\) and masses
\(m_1,\ldots,m_N\). The total mass of the system is
\[
M=\sum_{i=1}^{N}m_i.
\]

The microscopic mass-density field is defined as
\[
\rho(\mathbf{r},t)
=
\sum_{i=1}^{N}
m_i\,\delta\!\left(\mathbf{r}-\mathbf{r}_i(t)\right).
\]

Hence, provided all particles are contained within the domain \(\Omega\),
\[
\int_{\Omega}\rho(\mathbf{r},t)\,d\mathbf{r}
=
\sum_{i=1}^{N}m_i
=
M.
\]

Defining
\[
p(\mathbf{r};t)\equiv\frac{\rho(\mathbf{r},t)}{M},
\]
we immediately obtain
\[
\int_{\Omega}p(\mathbf{r};t)\,d\mathbf{r}=1.
\]

To establish that \(p(\mathbf{r};t)\) is the one-particle spatial probability
density, consider a particle \(I\) selected randomly with mass-weighted
probability
\[
P(I=i)=\frac{m_i}{M}.
\]
Let \(\mathbf{R}(t)=\mathbf{r}_I(t)\) denote the position of the selected
particle. For any measurable region \(A\subseteq\Omega\),

\[
\begin{aligned}
\text{P}\!\left(\mathbf{R}(t)\in A\right)
&=
\sum_{i=1}^{N}
\frac{m_i}{M}
\mathbf{1}_{A}\!\left(\mathbf{r}_i(t)\right)\\
&=
\frac{1}{M}
\sum_{i=1}^{N}
m_i
\int_A
\delta\!\left(\mathbf{r}-\mathbf{r}_i(t)\right)\,d\mathbf{r}\\
&=
\int_A
\frac{1}{M}
\sum_{i=1}^{N}
m_i
\delta\!\left(\mathbf{r}-\mathbf{r}_i(t)\right)\,d\mathbf{r}\\
&=
\int_A p(\mathbf{r};t)\,d\mathbf{r}.
\end{aligned}
\]

Therefore, by the definition of a probability density,
\[
p(\mathbf{r};t)=\frac{\rho(\mathbf{r},t)}{M}
\]
is precisely the one-particle spatial probability density associated with
the mass-density field \(\rho(\mathbf{r},t)\).

For identical particles, \(m_i=m\) and \(M=Nm\), and the result reduces to
\[
p(\mathbf{r};t)
=
\frac{1}{N}
\sum_{i=1}^{N}
\delta\!\left(\mathbf{r}-\mathbf{r}_i(t)\right),
\]
which is the usual empirical one-particle spatial density.

\section{Connection Between Spatial Shannon Entropy and Gibbs Entropy}
\label{app:connection}

The Shannon entropy considered in this work is defined from the one-particle spatial probability density $p(\mathbf{r};t)$ and therefore represents a spatial differential entropy. It is important to distinguish this quantity from the Gibbs entropy, which is defined over the full phase space of a statistical-mechanical system. For a classical system described by the position and momentum random variables $X$ and $P$, the differential Shannon entropy satisfies the chain rule
\begin{equation}
H(X,P)=H(X)+H(P|X).
\end{equation}

Consequently,
\begin{equation}
\frac{dH(X,P)}{dt}=\frac{dH(X)}{dt}+\frac{dH(P|X)}{dt}.
\end{equation}

Consider now a system satisfying the following assumptions: (i) the particle composition and particle masses are fixed; (ii) the system is maintained at a constant temperature; (iii) the kinetic and configurational degrees of freedom are separable, such that the Hamiltonian may be written as
\begin{equation}
\mathcal{H}(\mathbf{x},\mathbf{p})=K(\mathbf{p})+U(\mathbf{x}),
\end{equation}
with the kinetic energy independent of the spatial coordinates; and (iv) the momentum distribution remains the corresponding equilibrium distribution determined by the fixed masses and temperature. Under these conditions, the conditional momentum entropy is time-independent,
\begin{equation}
\frac{dH(P|X)}{dt}=0,
\end{equation}
and hence
\begin{equation}
\frac{dH(X,P)}{dt}=\frac{dH(X)}{dt}.
\end{equation}

In classical statistical thermodynamics, the Gibbs entropy is related to the phase-space differential entropy, apart from the conventional dimensional constant associated with the phase-space measure, by a factor of Boltzmann's constant. Thus, under the assumptions above, and when the spatial entropy relevant to the thermodynamic description is represented by the spatial probability density considered in this work,
\begin{equation}
\frac{dS_{\mathrm{G}}}{dt}=k_{\mathrm{B}}\frac{dH}{dt}.
\end{equation}
This establishes a correspondence between the temporal evolution of the spatial Shannon entropy used in the present framework and the thermodynamic Gibbs entropy under the stated assumptions \cite{HnizdoGilson2010}.

Accordingly, under the stated assumptions, the second law provides a thermodynamic interpretation of the entropy evolution derived in this work: the sign of $dH/dt$ is directly related to the sign of the corresponding thermodynamic entropy change through $dS_{\mathrm{G}}/dt=k_{\mathrm{B}}\,dH/dt$. Thus, the spatial Shannon entropy employed here may be regarded as a thermodynamically relevant entropy measure for the class of systems satisfying these assumptions, rather than as a general replacement for Gibbs entropy.
  
\end{document}